\documentclass[a4paper,10pt,twocolumn]{article}

\usepackage[english]{babel}
\usepackage[utf8]{inputenc}
\usepackage[T1]{fontenc}
\usepackage{url}

\usepackage{dsfont}
\usepackage{algorithm}
\usepackage{algorithmic}

\usepackage[top=1.5cm, left=1.5cm, right=1.5cm, bottom=1.5cm]{geometry}

\renewenvironment{abstract}{\bf\small {\em\ Abstract---}}{}

\usepackage{amsfonts,amssymb,amsmath,amsthm}
\usepackage{subfigure}
\usepackage{graphicx}
\usepackage[footnotesize]{caption}

\usepackage{amsfonts,amssymb,amsmath,amsthm}

\title{Phase inpainting in time-frequency plane}

\author{A.~Marina Kr\'em\'e $^{1,2}$\thanks{
This work was supported by the Agence Nationale de la Recherche under grant JCJC MAD (ANR-14-CE27-0002).}, Valentin Emiya$^2$ and Caroline Chaux $^{1}$\\
  \footnotesize $^1$Aix Marseille Univ, CNRS, Centrale Marseille, I2M,  Marseille, France\\ 
     \footnotesize $^2$Aix Marseille Univ, Universit\'e de Toulon, CNRS, LIS, Marseille, France\\ 
 }
  \date{\empty} 

\newcommand{\stft}{\ensuremath{\operatorname{STFT}}}
\newcommand{\istft}{\ensuremath{\operatorname{STFT}^{-1}}}
\DeclareMathOperator*{\Diag}{Diag}
\DeclareMathOperator*{\Trace}{Trace}
\DeclareMathOperator*{\Rang}{Rank}

\def\C{\mathbb{C}}

\newcommand{\dotprod}[2]{\left\langle #1, #2 \right\rangle}

\def\sig{\mathbf{x}}
\def\sigrec{\mathbf{\hat{x}}}
\def\siglen{N} 
\def\projvec{\mathbf{a}} 
\def\projmat{\mathbf{A}} 
\def\measvec{\mathbf{y}}
\def\knampvec{\mathbf{b}} 
\def\bmaskvec{\boldsymbol{m}} 
\DeclareMathOperator*{\support}{supp} 
\def\phaseamp{\mathbf{b}}

\def\liftsig{\mathbf{X}} 
\def\phasevec{\mathbf{u}} 
\def\phasemat{\mathbf{U}} 
\def\liftproj{\mathbf{A}} 

\def\nbbins{F}
\def\nbframes{T}
\def\winvec{\mathbf{w}}

\begin{document}

\maketitle

\begin{abstract} 
We propose a new problem of missing data reconstruction in the time-frequency plane. This problem called phase inpainting, consists in reconstructing a signal from time-frequency observations where all amplitudes and some phases are known while the remaining phases are missing.  A mathematical formulation of this problem is given. We propose three alternatives of existing algorithms. An iterative algorithm: Griffin and Lim and two  semidefinite programming optimization algorithms: PhaseLift and PhaseCut.
The obtained results show that knowledge of certain phases  improves the reconstruction's quality. 

\end{abstract}

\section{Introduction}
\label{sec:introduction}

Reconstructing missing data in the time-frequency plane is a difficult problem that may be addressed in two successive steps. First, the missing amplitudes are reconstructed, as in ~\cite{Smaragdis2009,LeRoux2010,Hamon2016}. Then, we may call \textit{phase inpainting} the remaining task in which all amplitudes and some phases are known while the remaining missing phases must be reconstructed. We propose to formulate and address this phase inpainting problem, which has not been addressed so far and remains a challenge.
The organization of this paper  is as follows. In section \ref{sec:first-section}, the problem of phase inpainting is formalised. In sections \ref{sec:GLI}, \ref{sec:PLI}, and \ref{sec:PCI} we present our Griffin and Lim algorithm for phase inpainting (GLI), PhaseLift for phase inpainting (PLI), and PhaseCut for phase inpainting (PCI). In Section 6, some small experiments with different missing data ratios and several mask shapes illustrate their performance and limitations. Finally, conclusion and perspectives are drawn in Section 7.
Note that more details about the proposed work have been published in ~\cite{Kreme2018}.

\section{Phase inpainting problem}
\label{sec:first-section}

For $t \in \{0, \ldots, \nbframes-1\}$ and  $ \nu \in \{0, \ldots, \nbbins-1\}$, we define Gabor atoms by  $\projvec_{t, \nu}= \winvec[n-t h] e^{2\imath \pi \frac{\nu}{\nbbins} n}$ where 
$\winvec$  is the window  and $h$   the hop size. For a signal $\sig \in \C^\siglen$, we have  $\nbbins \times  \nbframes$  complex linear measurements $\phaseamp=\left [\dotprod{\sig}{\projvec_{t,\nu}}\right ]_{t,\nu=1}^{\nbbins \times \nbframes }\in\C^{\nbbins \times  \nbframes }.$ The short-time Fourier transform(STFT) is define by: $\stft_{x}[t,\nu]=\dotprod{\sig}{\projvec_{t, \nu}}= \projvec_{t, \nu}^H\sig$.We assume that we observe both the magnitudes and the phases of a subset of measurements while only the magnitudes of the remaining measurements is available. The location of these subsets is given by a known  binary mask $\bmaskvec\in\left \lbrace 0, 1\right \rbrace^ {\nbbins \times \nbframes}$ : $\bmaskvec\left [t,\nu \right ] = 1$ if both the magnitude and the phase of measurement $b[t,\nu]$ are known and $\bmaskvec\left [t,\nu\right ] = 0$ if only its magnitude is known.Then the phase inpainting problem is given by

\begin{small}
\begin{align}
\label{eqn:phase_inpainting}
\text{Find } \sig \in \C^\siglen \text{ s.t. } 
\begin{cases}
\dotprod{\sig}{\projvec_{t, \nu}}& =\phaseamp[t, \nu], \forall t, \nu \in \support\left (\bmaskvec\right )\\
\left | \dotprod {\sig}{\projvec_{t, \nu}}\right | & =\knampvec[t, \nu], \forall t, \nu \in  \support\left (\neg\bmaskvec\right ) 
\end{cases}
\end{align}
\end{small}

\section{Griffin and Lim for phase inpainting}
\label{sec:GLI}
A first approach we propose to solve problem ~\eqref{eqn:phase_inpainting} is an extension of the Griffin and Lim algorithm ~\cite{Griffin1984Signal}. The difference with the original is that we take into account the known phases. We denote by $\circ$ , the product of Hadamard, and  $\angle$ the angle of a complex number.

\begin{algorithm}[H]
\caption{Griffin and Lim for  phase inpainting (GLI)}
\begin{algorithmic}
\REQUIRE   
$ \begin{cases}
\phaseamp:  \text{observations} \; , 
\bmaskvec: \text{binary mask}\\
n_\text{iter}  : \text{number of iterations}\\
\stft \text{and} \istft:  \text{operators related to } \{a_{t,\nu}\}
 \end{cases}$
\STATE
\STATE Random initialization $\boldsymbol{\varphi}_{0}$ of  missing phases: $$  \boldsymbol{\varphi} \leftarrow \bmaskvec \circ \angle \phaseamp + (1-\bmaskvec)\circ \boldsymbol{\varphi}_0 \; \quad \text{and}\; \quad  \measvec^{(0)} \leftarrow \knampvec \circ \exp{(\imath \boldsymbol{\varphi})}   $$\\
\FOR {$i \in\left \lbrace 1,2,\ldots, n_\text{iter} \right \rbrace$}
 \STATE  $\mathbf{z}^{(i)}\leftarrow \stft \left( \istft \left (\measvec^{(i-1)} \right) \right)  $ \\
\STATE $\boldsymbol{\varphi}^{(i)}\leftarrow \bmaskvec \circ \angle \phaseamp + (1-\bmaskvec) \circ \angle{\mathbf{z}^{(i)}}$ 
\STATE  $\measvec^{(i)} \leftarrow \knampvec \circ \exp(\imath \boldsymbol{\varphi}^{(i)}) $
\ENDFOR
\RETURN $ \istft(\measvec^{(n_\text{iter})})$
\end{algorithmic}
\end{algorithm}

\section{PhaseLift for phase inpainting (PLI)}
\label{sec:PLI}

PhaseLift for phase inpainting (PLI) approach based on lifting and positive semidefinite programming (SDP) that we propose here is a variant of PhaseLift ~\cite{Candes2015a}. 


\label{prop:PLP}
With notations of problem~\eqref{eqn:phase_inpainting}, let $\liftproj_{(t,\nu),(t',\nu')}=\projvec_{t,\nu}\projvec_{t',\nu'}^{H}$ for $(t,\nu),( t',\nu')  \in  {\lbrace 0, \ldots,  \nbframes -1 \rbrace}\times{ \lbrace  0,\ldots, \nbbins-1  \rbrace }$. Using the lifting $\liftsig = \sig \sig^{H}$, problem \eqref{eqn:phase_inpainting} is equivalent to: 

\begin{footnotesize}
\begin{equation}
\label{eqn:phaseliftrank}
\begin{aligned}
&{\min_{\liftsig \in \mathbb{C}^{\siglen \times \siglen}}} {\Rang ( \liftsig)}\\
  &\mathrm{ s.t. } 
\begin{cases}
\Trace ( \liftproj_{ (t,\nu), (t',\nu')}\liftsig) = \phaseamp[t,\nu] \bar{ \phaseamp}[t',\nu'],\;  \forall \; (t,\nu), (t',\nu') \in \support (\bmaskvec) \\ 
\Trace ( \liftproj_{(t,\nu),(t,\nu)}\liftsig) =\knampvec^2[t,\nu],  \; \forall \;  (t,\nu) \in \support (\neg\bmaskvec)\\ 
\liftsig\succeq 0
\end{cases}
\end{aligned}
\end{equation}
\end{footnotesize}

and can be relaxed as :
\begin{footnotesize}
\begin{equation}
\begin{aligned}
\label{eqn:phaselifttrace}
&{\min_{\liftsig \in \mathbb{C}^{\siglen \times \siglen}}} {\Trace ( \liftsig)}\\
  &\mathrm{ s.t. } 
\begin{cases}
\Trace ( \liftproj_{ (t,\nu), (t',\nu')}\liftsig) =  \phaseamp[t,\nu] \bar{ \phaseamp}[t',\nu'], \; \forall \; (t,\nu),(t',\nu') \in \support (\bmaskvec) \\ 
\Trace ( \liftproj_{(t,\nu)(t,\nu)}\liftsig) =\knampvec^2[t,\nu], \forall  (t,\nu) \in \support (\neg\bmaskvec) \\ 
\liftsig\succeq 0
\end{cases}
\end{aligned}
\end{equation}
\end{footnotesize}
PLI differs from PhaseLift ~\cite{Candes2015a} in the additionnal constraints (first row in ~\ref{eqn:phaselifttrace}) where known phases appear.

\section{PhaseCut for phase inpainting (PCI)}
\label{sec:PCI}
PhaseCut for phase inpainting (PCI) is also an SDP optimization algorithm. It is a variant 
of the original PhaseCut ~\cite{Waldspurger2015}. The idea of this method is to split 
%
 the amplitude and phase variables, so that one may optimize only on the phase  vector 	$\phasevec\in\C^{\nbbins \nbframes}$ such that $\forall \;  k, \vert \phasevec\left [k\right ]\vert = 1 $. Using  the lifting $\phasemat = \phasevec \phasevec^H$ and the  notation of problem~\eqref{eqn:phase_inpainting}, let $ \boldsymbol{\Gamma}=\Diag(\mathbf{c}^{H})(I- \projmat \projmat^{\dagger} )\Diag(\mathbf{c})$,
$ \mathbf{c}\in\C^{\nbbins  \nbframes}$ is defined by 
$\mathbf{c} [k]  = \left | \phaseamp[k] \right |, \forall  k$.
Then problem~\eqref{eqn:phase_inpainting} is equivalent to

\begin{footnotesize}
\begin{equation}
\begin{aligned}
\label{eqn:pcirank}
&\min_{ \phasemat \in \mathbb{C}^{{\nbbins  \nbframes }\times {\nbbins  \nbframes }}} \Trace(\phasemat \boldsymbol{\Gamma})\\
& \mathrm{s.t. }  
\begin{cases} 
   \Diag(\phasemat) =\mathds{1} \\
   \phasemat[(t, \nu),(t', \nu')]= \frac{\phaseamp[t, \nu]}{\vert \phaseamp[t, \nu]\vert} \frac{\bar{\phaseamp}[t', \nu']}{\vert \phaseamp[t', \nu']\vert} \; \forall  (t,\nu), (t',\nu') \in \support (\neg\bmaskvec)\\
   \Rang\left (\phasemat\right ) = 1\\
   \phasemat \succeq 0          
\end{cases}
\end{aligned}
\end{equation}
\end{footnotesize}
and may be relaxed into a convex problem by dropping the rank constraint as

\begin{footnotesize}
\begin{equation}
\begin{aligned}
\label{eqn:pci}
&\min_{ \phasemat \in \mathbb{C}^{{\nbbins  \nbframes }\times {\nbbins  \nbframes }}} \Trace(\phasemat \boldsymbol{\Gamma})\\
&\mathrm{ s.t. }  
\begin{cases} 
   \Diag(\phasemat) =\mathds{1} \\
      \phasemat[(t, \nu),(t', \nu')]= \frac{\phaseamp[t, \nu]}{\vert \phaseamp[t, \nu]\vert} \frac{\bar{\phaseamp}[t', \nu']}{\vert \phaseamp[t', \nu']\vert} \; \forall  (t,\nu), (t',\nu') \in \support (\neg\bmaskvec)\\
   \phasemat \succeq 0.      
\end{cases}
\end{aligned}
\end{equation}
\end{footnotesize}
As for  PLI, PCI shows phase difference constraints (second row in ~\ref{eqn:pci} ) which makes the difference with the original PhaseCut ~\cite{Waldspurger2015}.
\section{Experiments}
\label{sec:simul}

\begin{figure}[tbh]
\centering
\includegraphics[width=0.22\textwidth]{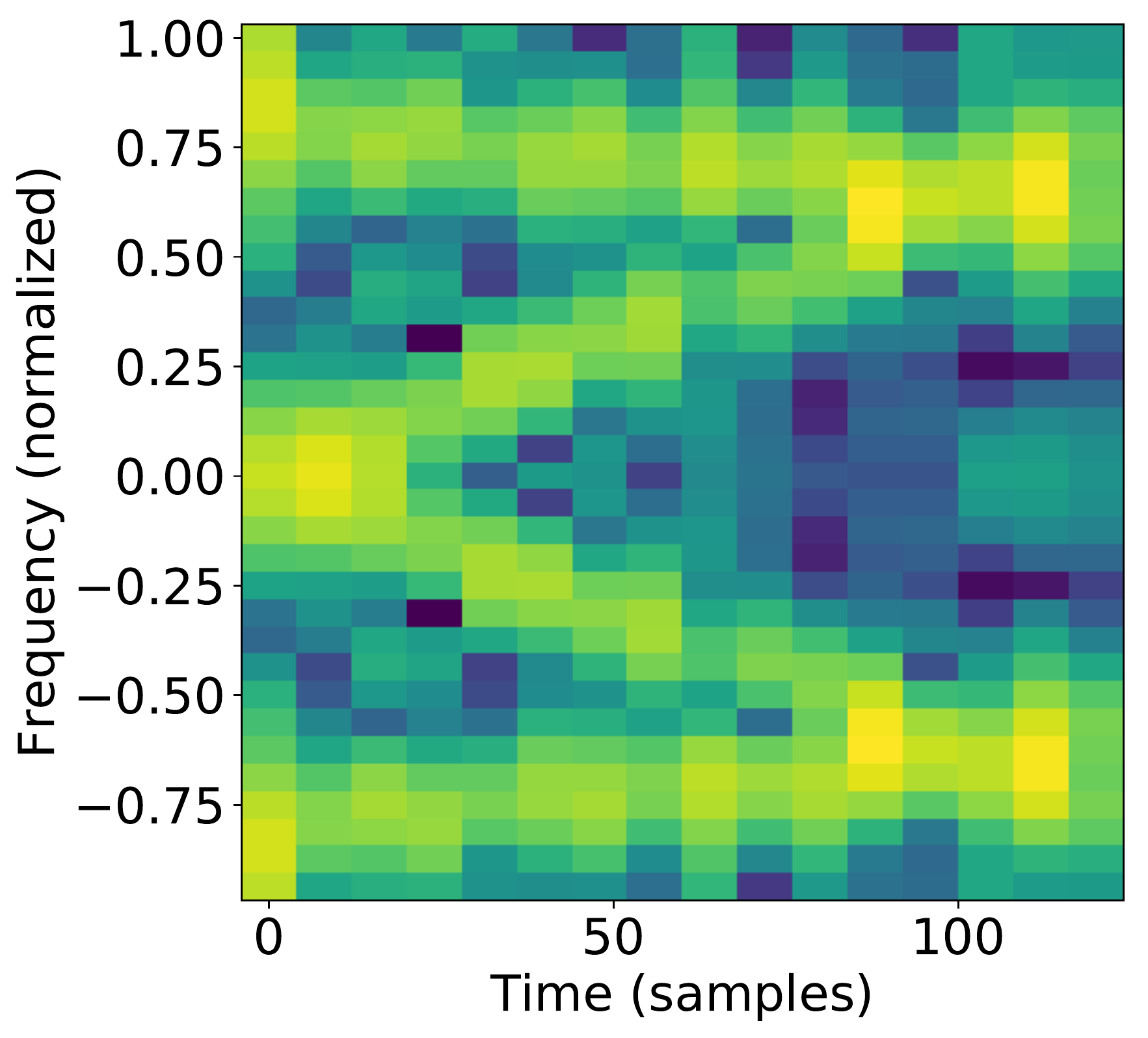} \hfill
\includegraphics[width=0.22\textwidth]{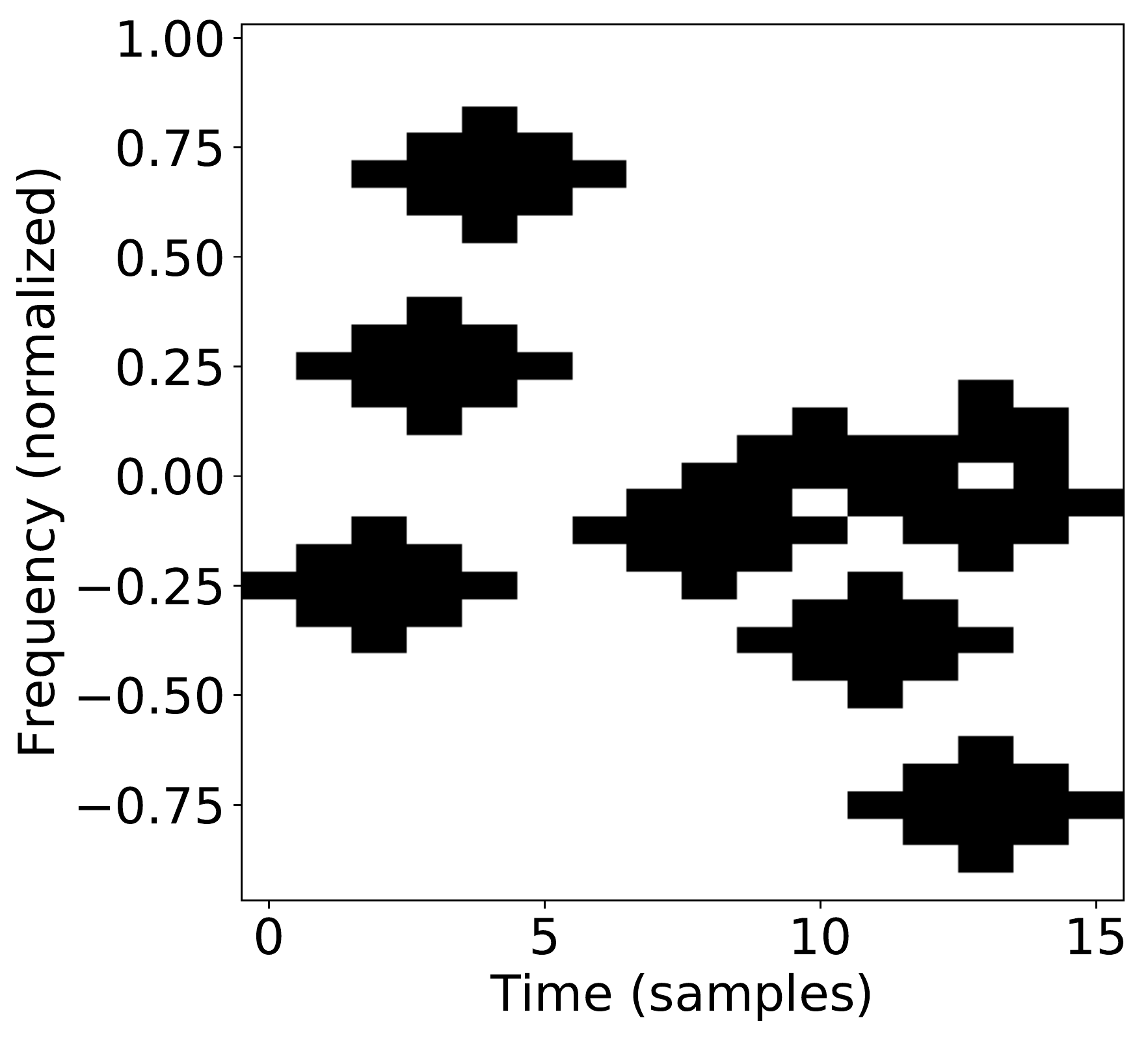} 
\caption{Spectrogram of the signal (left),$\nbframes = 16$ and $\nbbins=32$   and example of a mask with random holes of width 5 in black (right).}
\label{fig:data}
\end{figure}

We performed the experiments with a signal of length $\siglen=128$  composed of a mixture of two linear real chirps with normalized  frequency ranges$(0, 0.8)$ and $(0.8, 0.6)$, a dirac located at sample $64$ and a white Gaussian noise at a signal-to-noise ratio of 10 dB.
The STFT is obtained with a Hann window length  $16$, a hope size of $8$ samples (i.e., $\nbframes = 16$ frames  and $\nbbins=32$ frequency bins). While PLI implemenation was done using the TFOCS \cite{Becker2010} toolbox, PCI  implementation was done using  Block coordinate descent  \cite{Wen2012} methods.
Two experiments were carried out taking into account the type of masks and the percentage of missing data. In the first experiment, we consider a mask that randomly and uniformly generates the missing phases according to the percentage of missing data. In the second experiment, the percentage of missing data is fixed at  $30\%$ and the size of the holes varies from 1 to 9  coefficients. Figure~\ref{fig:data} illustrates the STFT of the signal and of one generated mask.
 

A baseline approach is also used, denoted as Random Phase Inpainting (RPI) and consisting in filling the missing phases by drawing random values independently and uniformely in $\left [0, 2\pi\right [$. We measure the performance between the original signal $ \sig$  and the reconstructed signal $ \sigrec$ up to a global phase  $ \theta \in \left [0, 2\pi\right [$ by 
$ E_{dB}(\sig,\sigrec) = 20\log_{10} \min_\theta \frac{{\Vert \sig- e^{\imath\theta}\sigrec \Vert }_{2}}{{\Vert \sig\Vert_{2}}}$ where $ \sig$ denotes the original signal and $ \sigrec$ the reconstructed one.

\begin{figure}[tbh]
\centering
\includegraphics[width=0.28\textwidth]{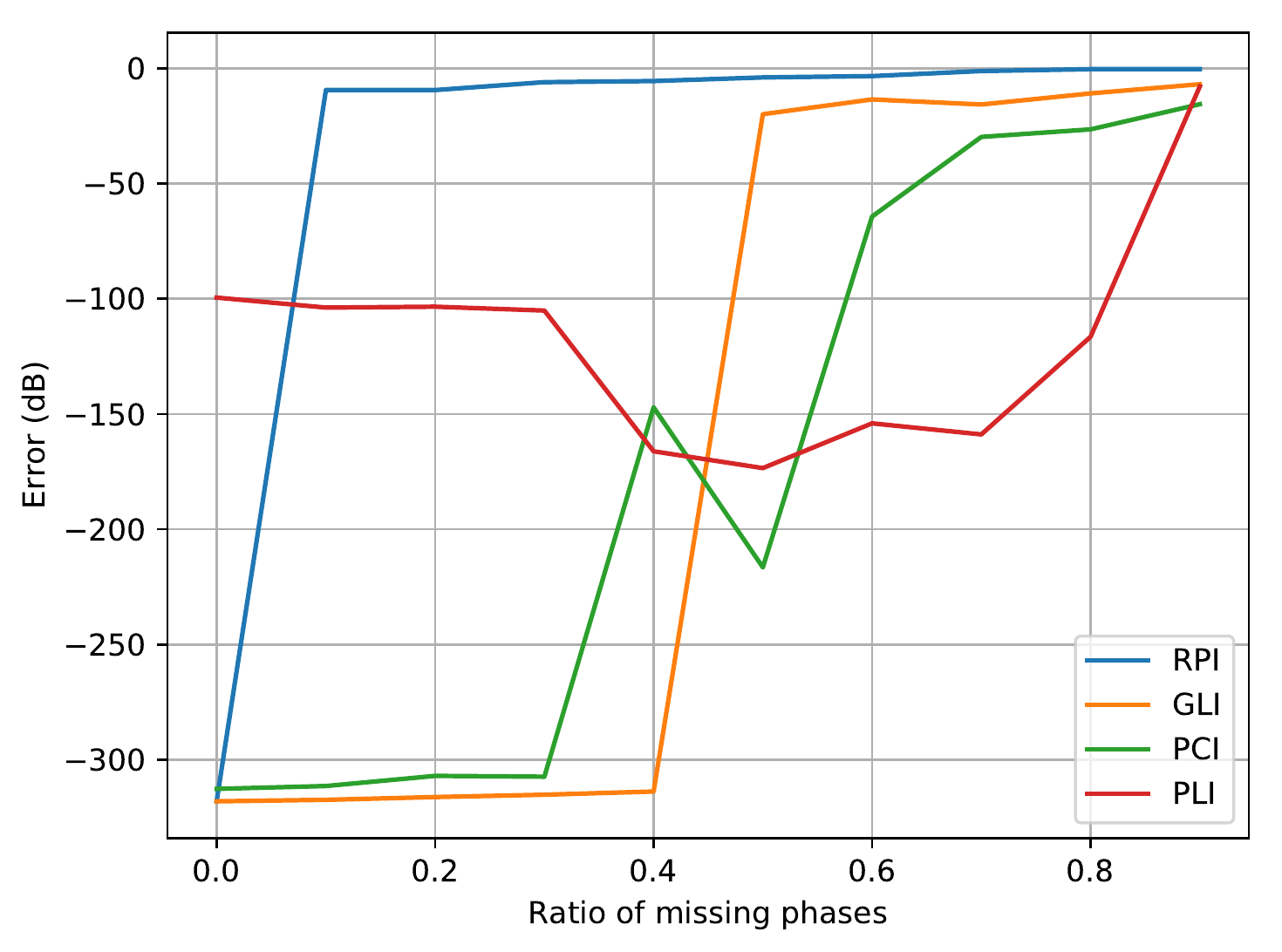}
\includegraphics[width=0.28\textwidth]{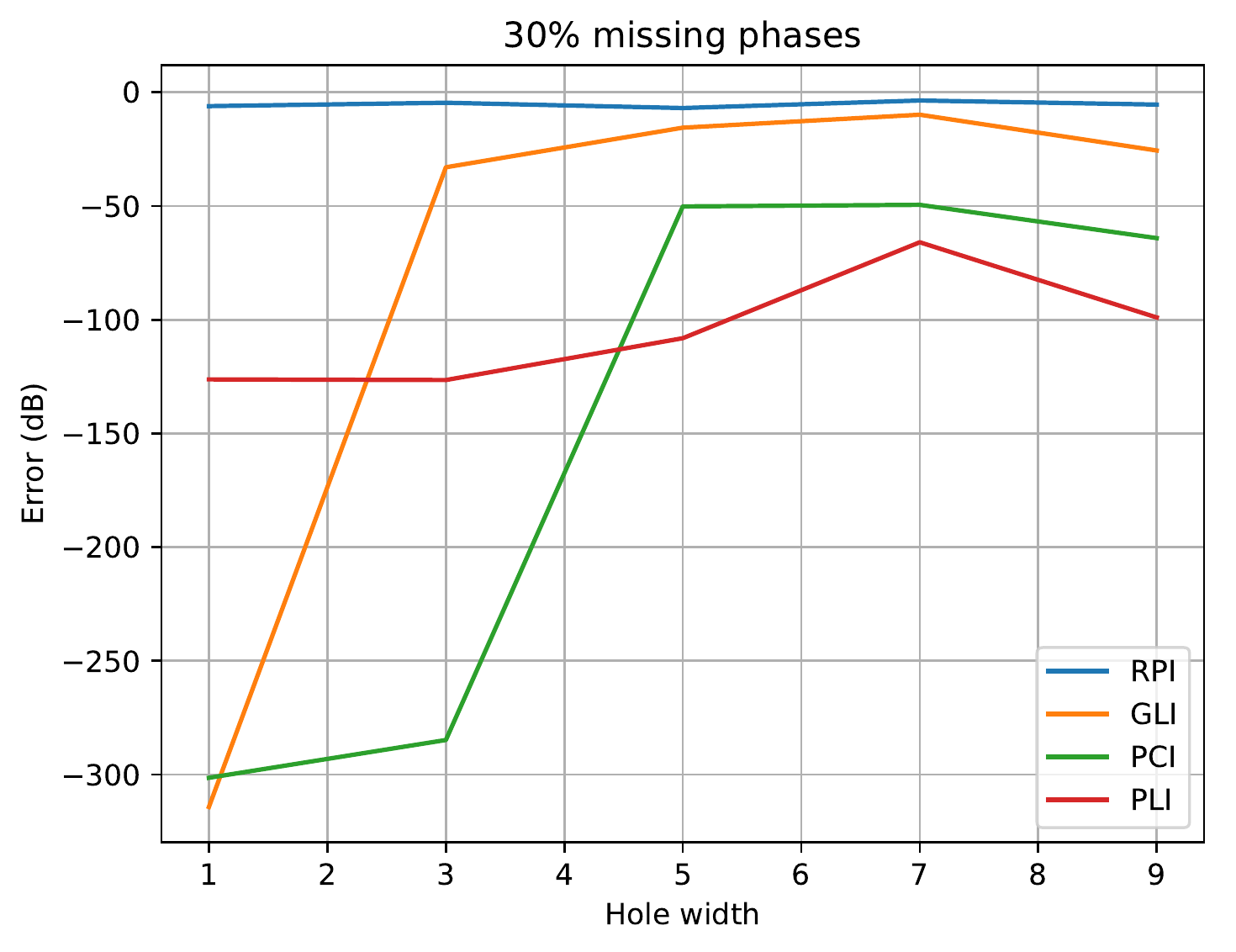} 
\caption{Reconstruction error as a function of the ratio of missing phases randomly distributed (top), for $30\%$ missing phases, as a function of the width of randomly distributed holes (bottom).}
\label{fig:erreur}
\end{figure}

In Figure~\ref{fig:erreur} (top), the reconstruction errors of the methods described in this paper are presented and compared with each other according to the percentage of missing data in the case of a random mask. We see that from $0$ to $30\%$, GLI and PCI achieve a perfect reconstruction unlike PLI which is good but not perfect. GLI remains perfect up to $40\%$ compared to PLI and PCI. From  $40\%$, PCI and PLI become better than GLI with a reconstruction error of less than -50dB. Figure ~\ref{fig:erreur} (bottom) shows the reconstruction errors as a function of hole width. As before, we see that the SDP methods are better than GLI which is only perfect for holes of width $1$.

%

PLI and PCI suffer from slowness due to lifting.  The PLI execution time is at least $ 6$ hours for one call when the number of missing phases is important. PCI convergence is from about 4 hours for $105$ iterations for a significant number of missing phases. GLI converges before the maximum number of iterations, with a runtime of less than one second for each call. More  details on the computational complexity and the convergence are given in \cite{Kreme2018}.



\section{Conclusion and perspectives}
\label{sec:conclu}
We have proposed a new problem of phase inpainting in time-frequency plane.Three algorithms including variants of the SDP algorithms have been proposed. Experiments with synthetic signals have shown that SDP methods (PLI, PCI) are better than alternate projections (GLI). In order to benefit from SDP results, one may investigate the adaptation of SDP algorithms to process only a local time-frequency region instead of the whole STFT matrix.Other algorithms may be designed for phase inpainting.In particular, some recent contributions to phase retrieval~\cite{Netrapalli2013,Candes2015Phase,Prusa2017,Dremeau2015,Metzler2017,Rajaei2017,Bahmani2016} may be adapted and may give good performance without the computational limits of SDP methods.

\bibliographystyle{plain}
\bibliography{biblio_stage.bib}
\end{document}